\documentclass[usenatbib,useAMS,aas_macros]{mn2e}
\usepackage{graphicx}
\usepackage{amssymb,amsmath}
\usepackage{xcolor}
\usepackage{url}    
\usepackage[
      colorlinks=false,    
      urlcolor=black,    
      citecolor=black,
      urlcolor=black,
      menucolor=black,    
      linkcolor=black,    
      bookmarks=true,    
      citebordercolor=lightgray,
      linkbordercolor=lightgray,
      urlbordercolor=lightgray,
      bookmarksopen=true,    
      hyperfootnotes=false,    
      pdfpagemode=UseOutlines    
]{hyperref}

\title[Modelling of the hard lags in GX 339--4]{Joint spectral-timing modelling of the hard lags in GX~339--4: constraints on reflection models}
\author[P. Cassatella et al.]{P. Cassatella$^{1,}$\thanks{E-mail: Pablo.Cassatella@soton.ac.uk}, P. Uttley$^{1,2}$, J. Wilms$^{3}$ and J. Poutanen$^{4}$\\
$^{1}$Astronomy Group, Faculty of Physical and Applied Sciences, University of Southampton, Southampton SO17 1BJ, United Kingdom\\
$^{2}$Astronomical Institute ``Anton Pannekoek", University of Amsterdam, Postbus 94249, 1090 GE Amsterdam, The Netherlands\\
$^{3}$Dr. Karl Remeis-Sternwarte \& ECAP, Universit\"at Erlangen-N\"urnberg, Sternwartstr. 7, 96049, Bamberg, Germany\\
$^{4}$Astronomy Division, Department of Physics, PO Box 3000, FI-90014 University of Oulu, Finland
}

\begin{document}
\date{Accepted 2012 February 21. Received 2012 February 21; in original form
2011 November 25}
\pagerange{\pageref{firstpage}--\pageref{lastpage}} \pubyear{2012}

\maketitle

\label{firstpage}

\begin{abstract}
The X-ray variations of hard state black hole X-ray binaries above
2~keV show `hard lags', in that the variations at harder energies
follow variations at softer energies, with a time-lag $\tau$
depending on frequency $\nu$ approximately as $\tau \propto \nu^{-0.7}$.
Several models have so far been proposed to explain this time delay,
including fluctuations propagating through an accretion flow, spectral variations during 
coronal flares, Comptonisation in the extended hot corona or a jet, 
or time-delays due to large-scale reflection from the accretion disc.
In principle these models can be used to predict the shape of the
energy spectrum as well as the frequency-dependence of the time-lags,
through the construction of energy-dependent response functions which
map the emission as a function of time-delay in the system. Here we use 
this approach to test a simple reflection model for the frequency-dependent
lags seen in the hard state of GX~339--4, by simultaneously fitting the model
to the frequency-dependent lags and energy spectrum measured by 
{\it XMM-Newton} in 2004 and 2009.  Our model cannot simultaneously fit
both the lag and spectral data, since the relatively large lags require
an extremely flared disc which subtends a large solid angle to the 
continuum at large radii, in disagreement with the observed Fe
K$\alpha$ emission.  Therefore, we consider it more likely that 
the lags $>2$~keV are caused by propagation effects in the accretion
flow, possibly related to the accretion disc fluctuations which have
been observed previously.  

\end{abstract}


\section{Introduction}
\label{sec:intro}

Black hole X-ray binaries (BHXRBs) have extensively been studied both in terms
of their spectra (where a soft, multi-colour disc black-body and a
harder power-law component play the main role) and in terms of their
variability, which can be straightforwardly quantified using the common
approach of computing the Power Spectral Density (PSD) of their
signal. In combination with the time-averaged X-ray spectrum, this 
approach has enabled the variability to be studied as a function of
spectral state, so that the hard power-law-dominated states can be
associated with large variability amplitudes (tens of per cent fractional
rms) and band-limited PSD shapes, while the soft disc-dominated states
show much weaker variability and broadband power-law like PSD shapes. 

Although they provide information on the global connection between the
emitting components and the variability process, the PSD and mean energy
spectra alone carry no information about the complex pattern of interlinkage 
between diverse regions in the accreting system that lead to transfer of
variability from one physical component to another over a range of 
time-scales.  In order to understand the causal connection between emission
processes that show different relative strengths in two different
energy bands, the time-lags between these two bands can be extracted
as a function of Fourier frequency.  To date, frequency-dependent lags
have been studied  mostly in the hard state of BHXRBs.  There the
lags ($\tau$) are 'hard', in the sense that variations in harder bands
lag behind variations in softer bands and depend on frequency as $\tau
\propto \nu^{-0.7}$, albeit with some sharper 'steps' in the
lag-frequency relation
\citep{miyamoto1988,cui97,nowak1999}. 
The general form of these lags has been explained by a variety of models,
including Comptonisation in the extended corona \citep{kazanas97} or 
a jet \citep{reig2003,kylafis2008}, spectral variability during coronal flares \citep{pf99}, 
accretion fluctuations propagating through a power-law
emitting region (e.g. a corona) with an energy-dependent radial
emissivity profile \citep{kotov2001,arevalo2006},  
or light-travel times to an extended reflecting region \citep{kotov2001,poutanen2002}.
Models invoking Comptonisation in the extended regions predict 
broader auto-correlation function for photons at higher  energies, which suffer
more scattering, which is opposite to what is observed \citep{maccarone2000,poutanen2001}.

Recent measurements of the
lag \citep{uttley2011} between variations of the accretion disc blackbody emission and
the power-law component strongly indicate that the variations are driven
by fluctuations propagating through the disc \citep{lyubarskii1997} to the power-law emitting
region, however this model cannot simply explain
the hard lags between bands where the power-law dominates.

Models to generate frequency-dependent lags work by determining the
response to an input signal of an emitting region which produces a hard
spectral component, e.g. a compact coronal region sandwiching the disc,
the upscattering region in the jet or an extended reflector.  The input
signal may be changes in the geometry, a fluctuation in accretion rate, seed photon illumination
or illuminating primary continuum.  The hard emitting region
can only respond after a delay which, broadly speaking, is set by the time
taken for the signal to propagate to and across the region.  The delay time
is determined by the signal speed (viscous time-scale or light-travel time)
and the size scale of the hard emitting region.  If the delay is large
compared to the variability time-scale, the variations of the hard emitting
region are smeared out and so the amplitude of variations of the lagging
component is reduced.  Thus the observed drop in lags with Fourier frequency
can be reproduced.

In principle, models for the emitting regions which can reproduce the
lags should also be able to reproduce the energy spectrum.  Therefore,
the combination of lag information with information on the X-ray
spectral shape should provide much greater constraints on models for
the emitting regions than either the commonly used spectral-fitting
methods or the rarely-attempted fits to timing data.  As a good
proof-of-principle, the simplest models to attempt this joint lag and
spectral-fitting are those where the lags are produced by reflection,
i.e. so-called `reverberation' lags.  Evidence for small (few to tens of
$R_{\rm g}$ light-crossing time) reverberation lags generated by reflection
close to the black hole has been seen in Active Galactic Nuclei
\citep{fabian2010,zoghbi2010,demarco2011,emmanoulopoulos2011} and BHXRBs
\citep{uttley2011}, however the much larger lags we consider here require
a larger scale reflector, perhaps from a warped or flaring outer-disc, which
was originally considered by \citet{poutanen2002}.  Reflection of hard photons
from a central corona off an accretion disc should produce both a reflection
signature in the spectrum and a frequency-dependent time-lag, the shapes of
which should depend on the exact geometry of the reflector as well as
on the location of the corona itself (for a central, point-like
corona, this would be the height above the disc). It has to be stressed
that only non-flat disc geometries are expected to contribute a significant
lag at low frequencies \citep{poutanen2002}.

The CCD technology of the EPIC-pn camera onboard {\it XMM-Newton}
provides a time resolution of 5.965 ms (frame readout time) in Timing
mode with a 99.7\% livetime and an energy resolution $\sim125$ eV at 6 keV
that combined, offer great potential for developing models that take
into account energy and variability information together down to millisecond
time-scales.  Our aim in the present paper is to fit the spectra and time
lags from an EPIC-pn Timing mode observation of a BHXRB simultaneously to 
discern whether reflection is the main driver of the
observed hard-to-medium lags. In order to achieve this, we have
developed \textsc{reflags}, a reflection model that assumes a flared accretion
disc acting as reflector (assuming the constant density ionised disc reflection
spectrum of \citealt{ballantyne2001}), and is able to
output for a given geometry either the resulting spectrum, or the expected lags as a
function of frequency. We use this model to fit
simultaneously the spectra and time-lags of the low-mass BHXRB GX 339--4 
in the {\it XMM-Newton} observations of the hard state obtained in 2004 
and 2009, and so determine whether the reflection model can explain the hard
lags observed in these observations while also remaining consistent with 
the X-ray spectrum. In principle, this combined approach could yield much
greater constraints on the outer disc geometry than can be obtained with
spectral-fitting alone, since low velocities of the outer regions of the 
disc cannot be resolved with CCD detectors, although these regions should
contribute significantly to the lags if they subtend a large solid angle
as seen from the source.

We describe our model in Section~\ref{sec:model} and the data reduction
and extraction of spectra and time-lags in Section~\ref{sec:datared}. 
In Section~\ref{sec:results} we discuss the results from fitting the energy
spectra and frequency-dependent time-lags of GX~339--4 together with our model,
and also compare the expected optical/UV reprocessing signature of the inferred
geometries with data from the {\it Swift} satellite.  We discuss our results and
the wider implications of our combined spectral-timing model-fitting approach
in Section~\ref{sec:discussion}.

\section{A flared accretion disc model}
\label{sec:model}

\subsection{Model parameters}

We consider a simple reflection model where the hard lags are produced 
by the light travel times from a variable power-law continuum source to the
surface of a flared accretion disc which absorbs and reprocesses the incident radiation 
or scatters it producing a hard reflection spectrum
\citep{george1991}.  Following the geometry
described by \citet{poutanen2002}, we have written \textsc{reflags},
an \textsc{xspec} model that can describe both the lags as a function of 
frequency and the mean spectrum, allowing
simultaneous fitting by tying together the parameters from the
spectral and lag fits.  We assume that the accretion disc is axially 
symmetric but has a height of the disc surface above the mid-plane
which depends on radius $r$ as a power-law 
\begin{equation}
z(r) = H_{\rm out}\left(\frac{r}{R_{\rm out}}\right)^\gamma ,
\end{equation}
where $H_{\rm out}$ is the height of the disc surface at the 
outermost radius $R_{\rm out}$ 
and $\gamma$ is the `flaring index'.

We place a point-like source of Comptonised
photons at a certain height $H_{\textrm{src}}$ above the disc, located in the axis of
symmetry of the system, and assume that the disc is truncated at some inner radius $R_{\rm in}$. 
Because the lags are proportional  to the distances, while  
the spectral distortions due to rapid rotations are a function 
of radii in units of gravitational radius $R_{\rm g}=GM/c^2$,  
we have to specify the black hole mass, which we assume to be $M_{\rm BH}=10 \mbox{M}_\odot$. 

The amount of reflection of Comptonised photons that is expected to
come from each region of the disc strongly depends on the
value $\gamma$, and the only case where the contribution of
outer radii to spectra and lags can be significant is for concave
($\gamma >1$) geometries \citep{poutanen2002}.

The model also depends on the two parameters that describe the spectral
shape: the incident power-law photon index
$\Gamma$ and the ionisation parameter $\xi$. Due to computational
constraints in the model evaluation, the latter is approximated to be
constant throughout the disc. For computational purposes,
the disc is also divided into 100 equally spaced azimuthal angles and
300 logarithmically spaced radii, to produce of a total of 30\ 000 cells.

The total energy-dependent power-law plus reflection luminosity per
unit solid angle at a time $t$ that is emitted by a system whose geometry is described
by the parameters above (represented by the set $\mathbf{\alpha}$)
and seen at an inclination angle $i$ with respect to the axis of
symmetry, equals:

\begin{align}
\begin{split}
L_{\rm tot}(E, t | \mathbf{\alpha},i) &= L_{\rm PL}(E, t) \\
& + a(E) \frac{\Omega_{\rm  eff}}{2\pi}\sum_l  L_{\rm PL}(E, t-\tau_l)
  \kappa_{\mathbf{\alpha},l}(E,i) , 
\end{split}
\label{eqn:refl}
\end{align}
where the sum is performed over the index $l$ that
represents a single disc cell, $L_{\rm PL}(E,t)$ is the luminosity per unit
solid angle produced by
power-law emission and $a(E)$ is the albedo function. 

The factor $\kappa_{\mathbf{\alpha},l}(E,i)$ contains the several projection
terms and solid angle corrections required for the $l$-th cell
for the geometry described by $\mathbf{\alpha}$ (see \citealt{poutanen2002}), as well as relativistic
Doppler and gravitational redshift corrections (following the
same simplified approach as the \textsc{diskline} model,
\citealt{fabian1989}), and is normalised so that
$\sum_l \kappa_{\mathbf{\alpha},l}(E,i) = 1$.
For a given geometry $\mathbf{\alpha}$ and inclination
angle $i$, $\Omega_{\rm eff}$ then equals the solid angle subtended by the
disc, corrected for the source inclination as seen by the observer.
Finally, $\tau_l$ represents the time delay between observed direct
power-law emission and reflected emission coming from the $l$-th
cell.

\subsection{System response and time-lags}

For a direct power-law emission pulse (a delta-function)
equation~(\ref{eqn:refl}) can be rewritten as:
\begin{align}
\begin{split}
L_{\rm tot}(E, t | \mathbf{\alpha}, i)& = L_{\rm PL}(E) \\ 
& \times \Big (\delta(t) + 
a(E)  \frac{\Omega_{\rm  eff}}{2\pi} \sum_l
\kappa_{\mathbf{\alpha},l}(E,i) \delta(t-\tau_l)  \Big ) , 
\end{split}
\label{eqn:response}
\end{align}
and can be factorised as
\begin{equation}
L_{\rm tot}(E, t | \mathbf{\alpha}, i) =  L_{\rm PL}(E)T(E, t | \mathbf{\alpha}, i) ,
\label{eqn:factorised}
\end{equation}
where
\begin{align}
T(E, t | \mathbf{\alpha}, i ) = 
  \delta(t) + 
  a(E)  \frac{\Omega_{\rm  eff}}{2\pi} \sum_l \kappa_{\mathbf{\alpha},l}(E,i)
  \delta(t-\tau_l) . 
\end{align}
The system response function $T(E, t | \mathbf{\alpha}, i )$ contains the
energy-dependent time redistribution of an input power-law luminosity as a
function of time in a geometry $\mathbf{\alpha}$, as observed at an
inclination $i$.  Provided that the energy spectrum is averaged over a time
significantly longer than the largest $\tau_l$, $\tau_{\rm max}$ (which
in this case is of order 10~s), the spectrum is given simply given by: 
${\rm const.} \times \int_{0}^{\tau_{\rm max}} T(E, t' |
\mathbf{\alpha}, i ) {\rm d}t'$.
By taking the Fourier transform of equation~(\ref{eqn:factorised}):
\begin{align}
\begin{split}
\tilde{L}_{\rm tot}(E, \nu, i)&
  = \mathcal{F} \big \{ L_{\rm PL}(E) T(E, t | \mathbf{\alpha}, i)\big \} \\
& = L_{\rm PL}(E)\tilde{T}(E, \nu | \mathbf{\alpha}, i) .
\end{split}
\label{eqn:transforms}
\end{align}

In general, the time-lags between two light curves $s(t)$ and $h(t)$ in the soft and hard bands can be
calculated by computing their Fourier
transforms $\tilde{S}(\nu)= \mathcal{F}[s(t)]$ and $\tilde{H}(\nu)=\mathcal{F}[h(t)]$ and forming the
complex-valued quantity $C(\nu) = \tilde{S}^{*}(\nu)\tilde{H}(\nu)$,
called the cross-spectrum (where the asterisk denotes complex conjugation). 
Its argument is the phase lag or phase difference between $\tilde{S}(\nu)$ and
$\tilde{H}(\nu)$ \citep{nowak1999}:
\begin{align}
\phi(\nu) = \arg\big[ C(\nu ) \big] = \arg \big[ \langle \tilde{S}^{*}(\nu )
\tilde{H}(\nu )\rangle\big ]
\end{align}
and therefore   
\begin{align}
\tau(\nu ) = \frac{\phi(\nu)}{2\pi\nu}
\end{align}
equals the frequency-dependent time-lag.

The transfer function  $\tilde{T}(E, \nu | \mathbf{\alpha}, i)$ is equivalent to
the response function $T(E, t |  \mathbf{\alpha}, i)$ in the
Fourier-frequency domain. It is then possible to take the expression above to compute the
cross-spectrum that will give the time-lags
caused by reflection between two broad energy bands $s$ and $h$
and obtain the lags as above (the notation has been simplified):
\begin{align}
C(\nu) = L_{\rm PL}(E_{\rm s})L_{\rm PL}(E_{\rm h})
\tilde{T}^{*}(E_{\rm s}, \nu)\tilde{T}(E_{\rm h}, \nu) , 
\end{align}
\begin{align}
\tau(\nu) = \frac{ \arg\Big [ \tilde{T}^{*}(E_{\rm s},
\nu)\tilde{T}(E_{\rm h}, \nu) \Big ]}{2\pi\nu} .
\end{align}

\section{Data reduction}
\label{sec:datared}
\subsection{Extraction of spectra, event files and instrumental response files}

GX 339--4 was observed using the {\it XMM-Newton}  X-ray satellite on 2004 March 16
and 2009 March 26  (ObsIds 0204730201 and 0605610201, respectively).
In the present work, we will concentrate on the data taken using the EPIC-pn
camera in Timing Mode, which allows for a
fast readout speed, as opposed to imaging modes. This is done by
collapsing all the positional information into one dimension and
shifting the electrons towards the readout nodes, one macropixel row at a
time, thus
allowing a fast readout with a frame time of 5.965 ms \citep{kuster2002}.

The raw Observation Data Files (ODFs) were obtained from the {\it XMM-Newton} 
Science Archive (XSA) and reduced using
the {\it XMM-Newton}  Science Analysis System version 10.0.0 tool
\texttt{epproc} using the most recent Current Calibration Files
(CCFs).

With the aid of \textsc{evselect}, the source event information was then
filtered by column (\texttt{RAWX in [31:45]}), and pattern (\texttt{PATTERN <= 4},
corresponding to only single and double events), and only time
intervals with a low, quiet background (where \texttt{PI in [10000:12000]})
were selected for the subsequent data analysis. The total exposure times are $\sim 164$ ks
and $\sim 39$ ks for the 2004 and 2009 observations, respectively.
The background spectrum was obtained from ObsId 0085680501 
between columns 10 -- 18 when the source was fainter so that the background is
not contaminated by the source (see \citet{done2010} for further details).
\texttt{rmfgen} and \texttt{arfgen} were used to obtain the instrumental
response files. To account for the effects of systematic uncertainties in
the instrumentation, an additional 1\% was added to the error bars.

In the present work, uncertainties in the estimation of the
parameters are quoted at the 90\% confidence level for one parameter
of interest.

\subsection{Extraction of the time-lags}
\label{sec:datared:tlags}

When measuring the lags for real, noisy data one needs to first take
the average of the cross-spectrum over many independent light curve
segments (and also adjacent frequency bins).  This is because observational
noise adds a component to the cross-spectrum which has a phase randomly
drawn from a uniform distribution (see e.g. \citealt{nowak1999}). 
By averaging over many independent measures of the cross-spectrum,
the contribution of these noise components can be largely cancelled
out (the residual error provides the uncertainty in the lag).  Therefore,
the time-lags $ \tau(\nu)$ were extracted from the argument of the cross spectrum
averaged over many segments of the light curve  as the exposure time
and the dropouts due to telemetry saturation permit.
Throughout this work we will take as the soft band the interval 2.0--3.5 keV and 
as a hard band the interval 4.0--10.0 keV.

We choose to have 8192 bins/segment
for a total duration of $\sim 24.43$ seconds per segment, 
giving 6690 segments for
2004 and 1582 segments for 2009. 
This will also
constrain the frequency ranges over which lags can be obtained. These
frequencies have also been rebinned logarithmically with a step of
$\Delta\ln \nu=0.14$
in order to improve the signal-to-noise per frequency bin.

\begin{figure}
\centering
\includegraphics[width=.47\textwidth]{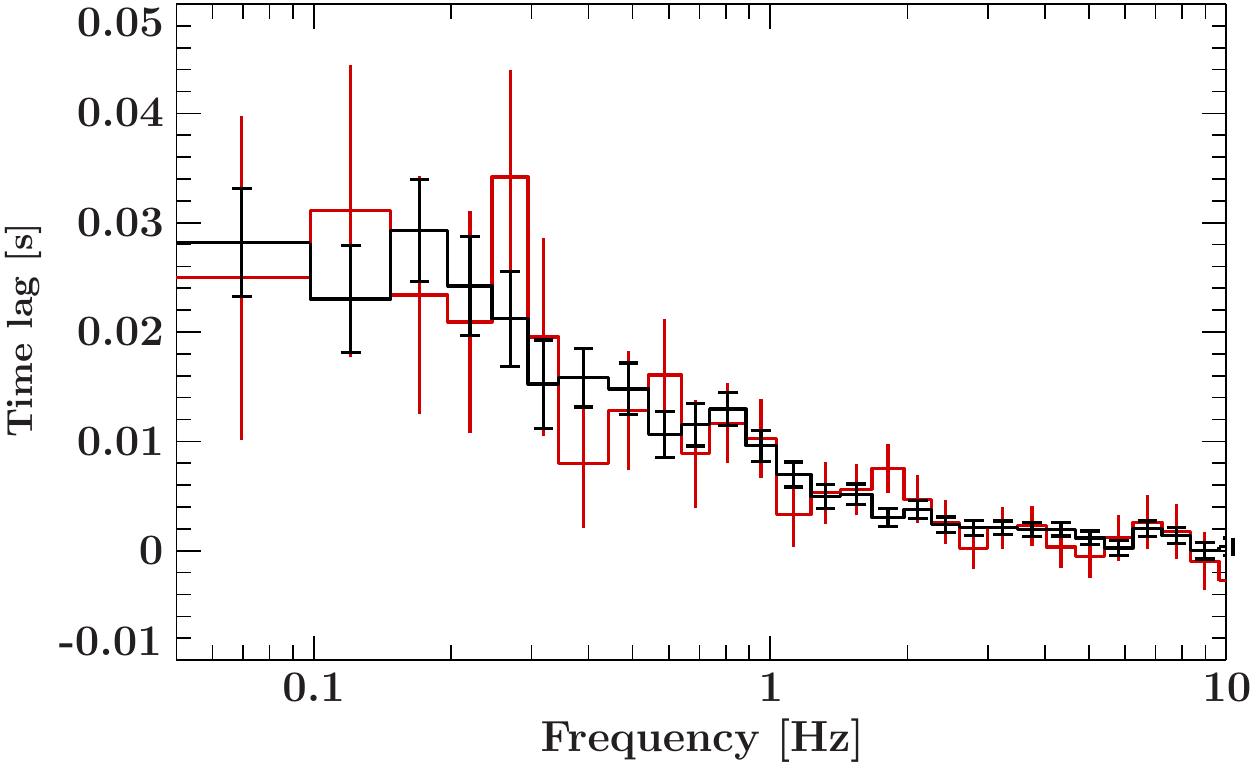}
\caption{
Time-lags between  4.0--10.0 and 2.0--3.5 keV energy bands 
as a function of frequency for the 2004 (black) and 2009
(red) observations. }
\label{fig:lagscomp}
\end{figure}

The uncertainties on the lag measurements follow from the technique in
\citet{nowak1999} (see also \citealt{bendat2010}). These scale as
$\sqrt{NM}$ where $N$ is the number of segments used to average the
cross-spectrum and $M$ is the number of frequencies averaged per bin. 
The lags are plotted on Fig.~\ref{fig:lagscomp}. 
Given the fact that, within the errors, the time-lag dependence on
Fourier frequency is consistent between the two observations, we
will sometimes assume that the time-lags for the two observations are
equivalent and do not vary between the two observations as their shapes
cannot be distinguished within the errors. This `substitution' should
in principle give us tighter constraints for our study in Section~\ref{sec:results}.

\section{Results}
\label{sec:results}

\begin{figure}
\centering
\includegraphics[width=0.47\textwidth]{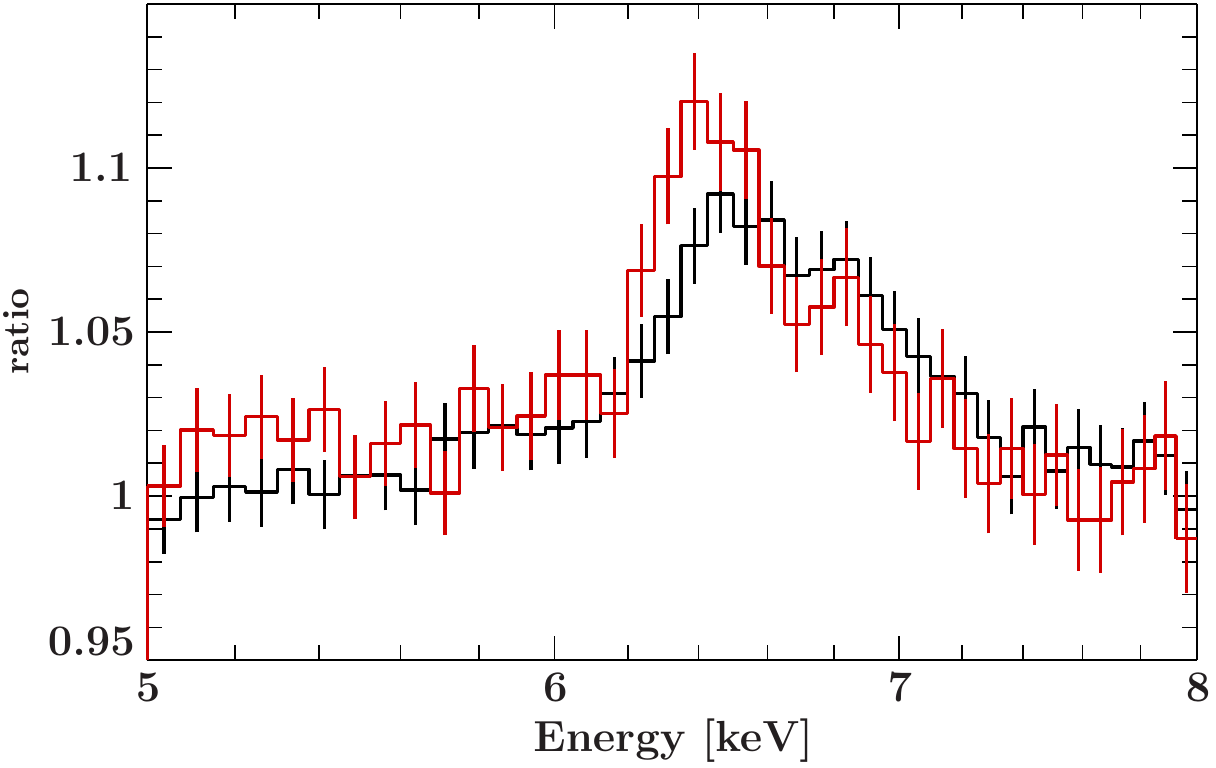}
\caption{Ratio to a powerlaw after fitting the 2.0--5.0 keV and 7.0--10.0 keV 
continuum (black: 2004, red: 2009). The presence of
the iron line is clear.}
\label{fig:rat}
\end{figure}

We test the model described in Section~\ref{sec:model} using the 2004 and
2009 observations of GX 339--4, in order to understand the validity of
the model and infer the geometrical parameters of the reflecting disc that
is required to explain both the spectrum and the lags.
To do this, the same model is concurrently fitted to the spectrum as well as the
lag versus frequency data.  Using this approach, it is possible to discern the importance of
reflection to explain the observed lags, and whether or not an extra
variability component to produce the lags is required.

In Fig.~\ref{fig:rat} we plot the ratio of the 2004 and 2009 spectra to
a power-law fitted to the 2.0--10.0 keV interval excluding the
region 5.0--7.0 keV. The iron line shape appears clearly different,
with the 2004 dataset showing a more broadened profile that is skewed towards higher energies, 
while the 2009 line
appears narrower and peaked at about 6.4 keV, the value expected for
neutral or weakly-ionised emission.

We perform fits using the 2004 and 2009 spectra in the 
3.0--10.0 keV band as well as their respective lags (we name these model fits  A and B,
respectively).  These energy intervals are chosen in order to avoid 
contamination from the band dominated by the disc blackbody emission. 
Due to the small uncertainties on the spectral data, the weak steepening
in spectral shape seen below 3~keV can skew the fit to the spectrum, hence we
cut off the spectral fit at 3 keV.  However, as shown by \citet{uttley2011},
the lags are not significantly affected by the disc at energies down to 2~keV,
hence we include photons down to this energy in the lag determination, in order
to increase signal-to-noise. 

\begin{table}
\centering
\small
{
\centering
\begin{tabular}{lcc}
\hline \hline
Parameter      & Obs. A  & Obs. B  \\ \hline
$R_{\rm in}/R_{\rm g}$  & $< 130 $ & $ < 220  $ \\ 
$R_{\rm out}/R_{\rm g}$ &$(16.8^{+2.4}_{-2.7}) \times 10^3$ & $(20^{+50}_{-8}) \times 10^3$ \\
$H_{\rm out}/R_{\rm out}$ &  $> 0.83$  & $> 0.46$  \\
$H_\textrm{src}/R_{\rm g}$ & $> 380 $ & $ > 330 $ \\
$\gamma$       &   $2.18^{+0.19}_{-0.24}$  & $> 1.69$ \\
$i$            &  $39.7^{+0.4}_{-0.5}$ &  $29.8^{+1.8}_{-4.3}$ \\
$\Gamma$ & $1.481^{+0.003}_{-0.002}  $ & $1.468 ^{+0.004}_{-0.006} $ \\
$\log \xi$          &   $2.05^{+0.03}_{-0.02}$   & $2.03^{+0.02}_{-0.01}$ \\
$\Omega^{\rm eff} / (2 \pi)$ & 1.13 & 1.28  \\ \hline
$\chi^2$ (spectrum)  &   1405  &   1305   \\
$\chi^2$ (time-lags)  &   152  &     32   \\
$\chi^2$/dof  &              1557 / 1417  &   1337 / 1416  \\
\hline
\end{tabular}
}
\caption{Spectral and lag fit parameters using the model
\textsc{reflags} in the band 3--10 keV. {
Hereinafter, we quote the total $\chi^2$ as well as the contribution of the $\chi^2$
that corresponds to both the spectrum and lags, separately.}}
\label{tbl:main}
\end{table}

\begin{figure*}
\centering
\includegraphics[width=1.\textwidth]{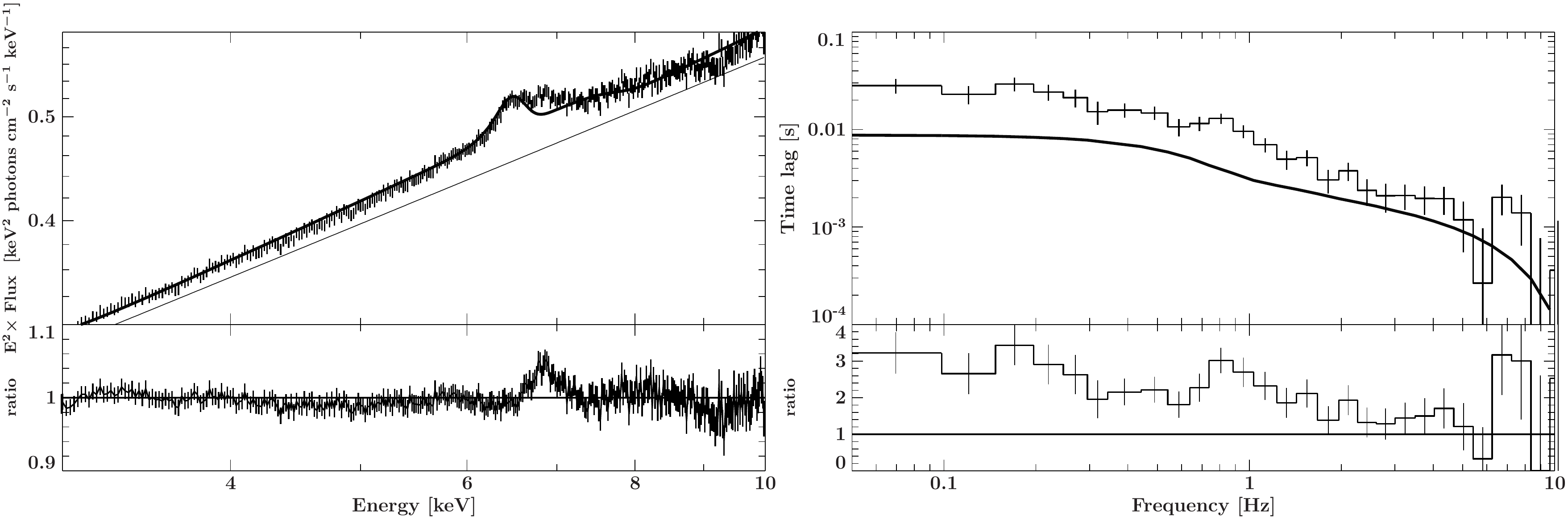}
\caption{Obs. A (2004) spectrum (left) and lags vs frequency (right). The thick solid
line represents the model for the best-fitting parameters (see
Table~\ref{tbl:main}, left column). {
In the spectrum, the thin solid line represents the
direct continuum power-law.} The lower
panels show the ratio data/model.}
\label{fig:04}                                    
\end{figure*}

\begin{figure*}
\centering
\includegraphics[width=1.\textwidth]{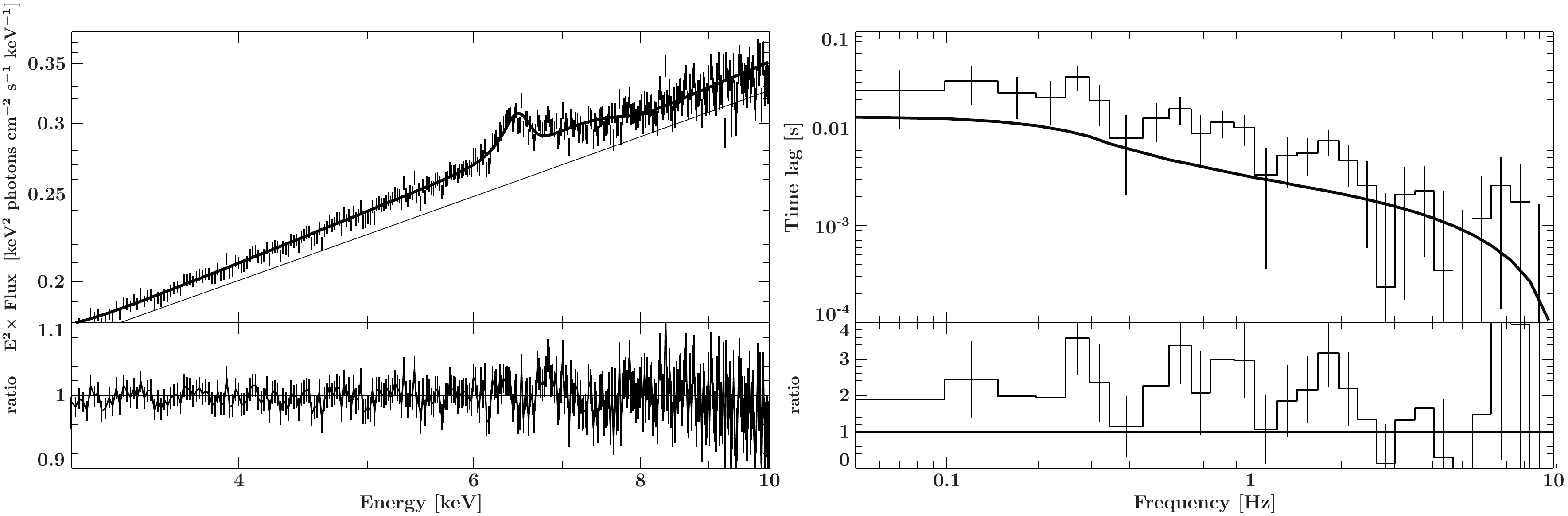}
\caption{Obs. B (2009) spectrum (left) and lags vs frequency (right). The thick solid
line represent the models {\sc reflags} for the best-fitting
parameters (see Table~\ref{tbl:main}, right column).}
\label{fig:09}                                    
\end{figure*}

The best-fitting parameters for fits A and B are shown in
Table~\ref{tbl:main}. The corresponding model comparisons to the data,
including the data-to-model ratios are plotted in Figs~\ref{fig:04} and \ref{fig:09}. 
These initial fits suggest that the model can represent the 2009 (B) data better. 
The upper limit on the disc inner radius is below 220 $R_{\rm g}$ in both cases. This value does not
give us any information on whether the accretion disc in GX 339--4 is 
truncated or not (for discussions about disc truncation, see e.g. \citealt{tomsick2009}).

As for the values of the outer radius, its value could reach up to 
$\sim 8\times 10^4$ $R_{\rm g}$ in B, much larger than what is found in A, indicating a
larger contribution to the narrow iron line in this case. This is in
agreement with the stronger core of the line in B as seen in the ratio
plot (Fig.~\ref{fig:rat}) as well as the difference between the solid
angles. An extremely high value for the source height $H_{\rm src}$
and a high value for the $H_{\rm out}/R_{\rm out}$ ratio suggest that the fits are
being driven by the lags, whose amplitude is strongly dependent on the
distances to the furthest regions of the reflector, even if their
contribution is small. The ionisation parameter remains consistent between
the observations, although a visual inspection of the residuals and a
clear difference in $\chi^2$ are indicative of a disc that can be
described with a line of ionised iron, and whose contribution
would likely come from the inner disc regions of the accretion disc where the
ionisation could plausibly be larger. 
{
The blue wing of a
relativistically broadened disc line could conceivably also contribute
to the residuals which we fit with a narrow line, however since the
inner disc radius is not strongly constrained by the lags, the model
is already relatively free to fit this feature with relativistic
emission, by being driven by the spectral data alone.  The fact that
it does not fit these residuals suggests that a more complex
ionisation structure of the disc is likely (also see
\citet{wilkinson2011}).}

\begin{figure*}
\centering
\includegraphics[width=1.\textwidth]{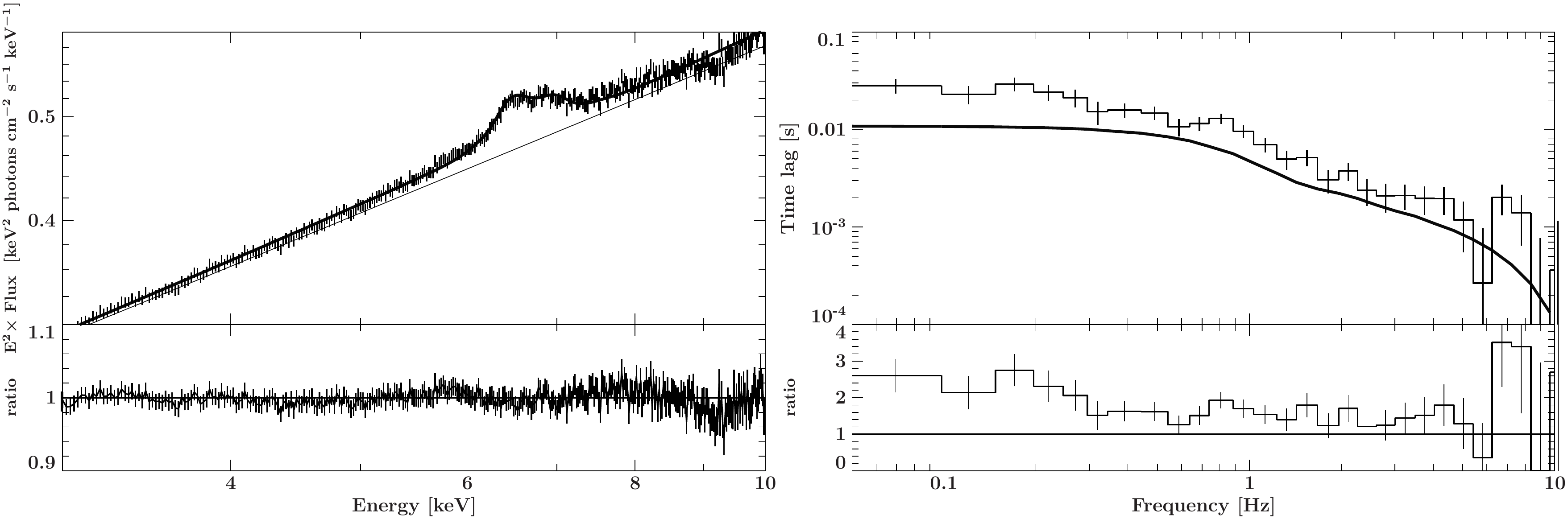}
\caption{Obs. A (2004) spectrum (left) and lags vs frequency (right). The thick solid
lines represent the model {\sc reflags} (including a Gaussian component in the
case of the spectrum) for the best-fitting parameters
(see Table~\ref{tbl:mix}, left column).}
\label{fig:04pl}                                    
\end{figure*}

\begin{figure*}
\centering
\includegraphics[width=1.\textwidth]{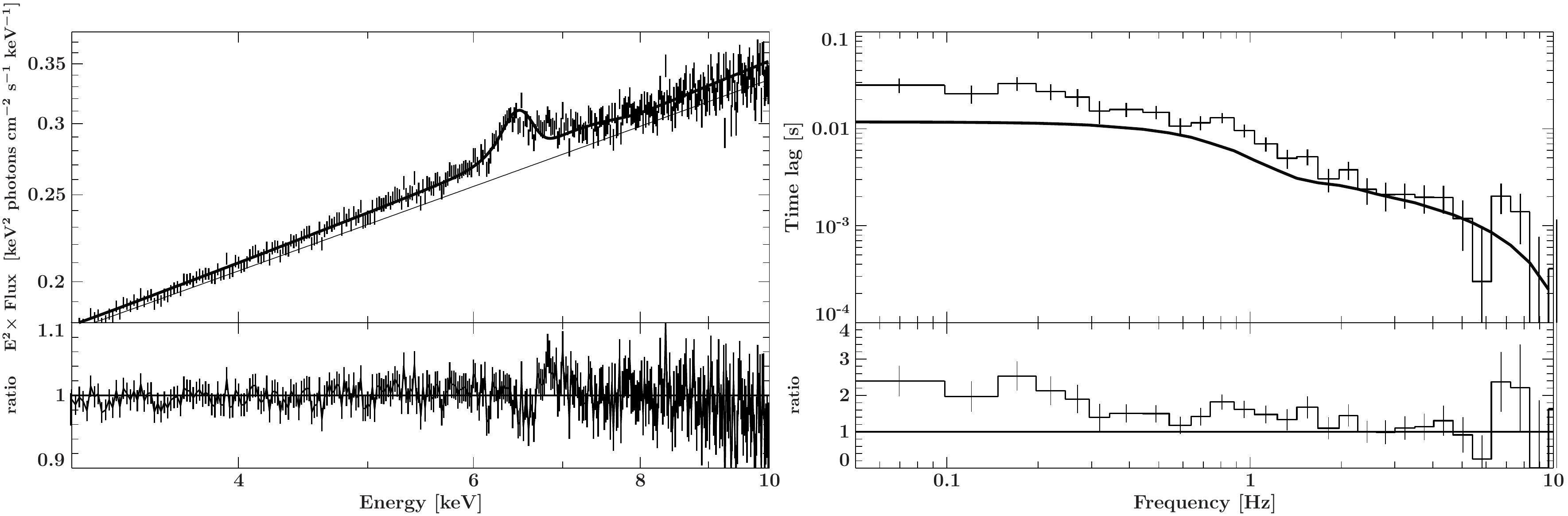}
\caption{Obs. B' (2009 spectrum, 2004 lags) spectrum (left) and lags vs frequency (right). The thick solid
lines represent the model {\sc reflags} for the best-fitting
parameters (see Table~\ref{tbl:mix}, right column).}
\label{fig:4+9}                                    
\end{figure*}

\begin{figure*}
\centering
\includegraphics[width=1.\textwidth]{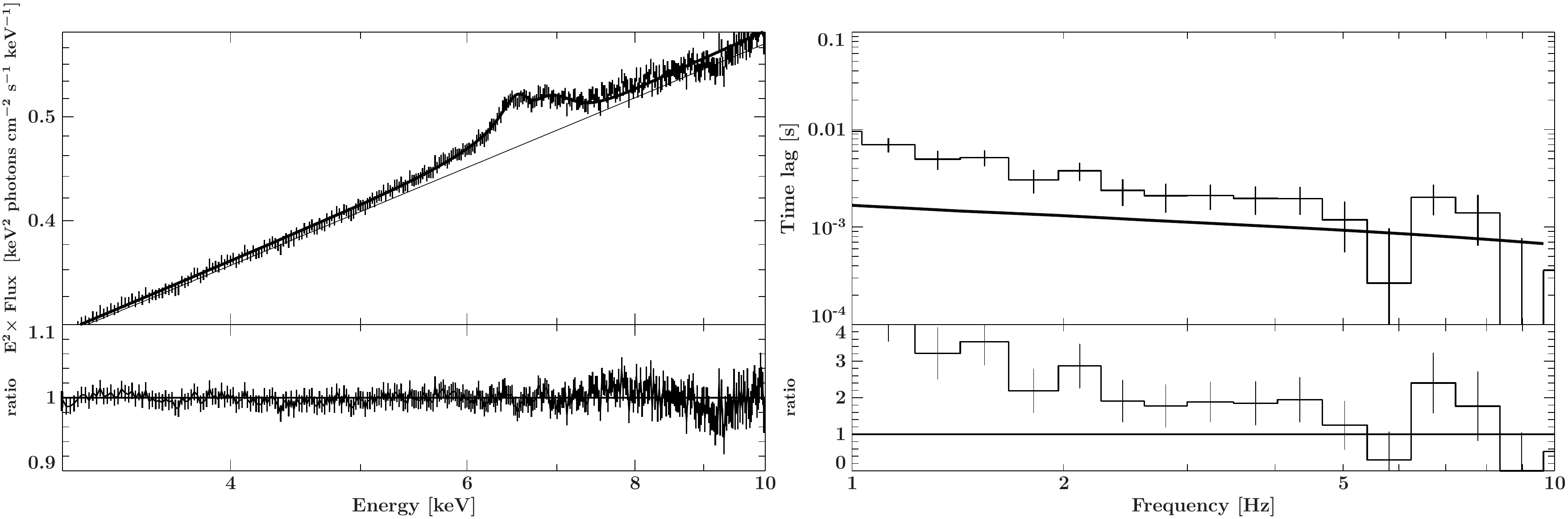}
\caption{Obs. A (2004 spectrum (left) and lags vs frequency (right). The thick solid
lines represent the model {\sc reflags} (including a Gaussian
component in the case of the spectrum) for the best-fitting
parameters (see Table~\ref{tbl:hifreq}, left column), in the frequency
range 1--10 Hz.}
\label{fig:04st}                                    
\end{figure*}

\begin{figure*}
\centering
\includegraphics[width=1.\textwidth]{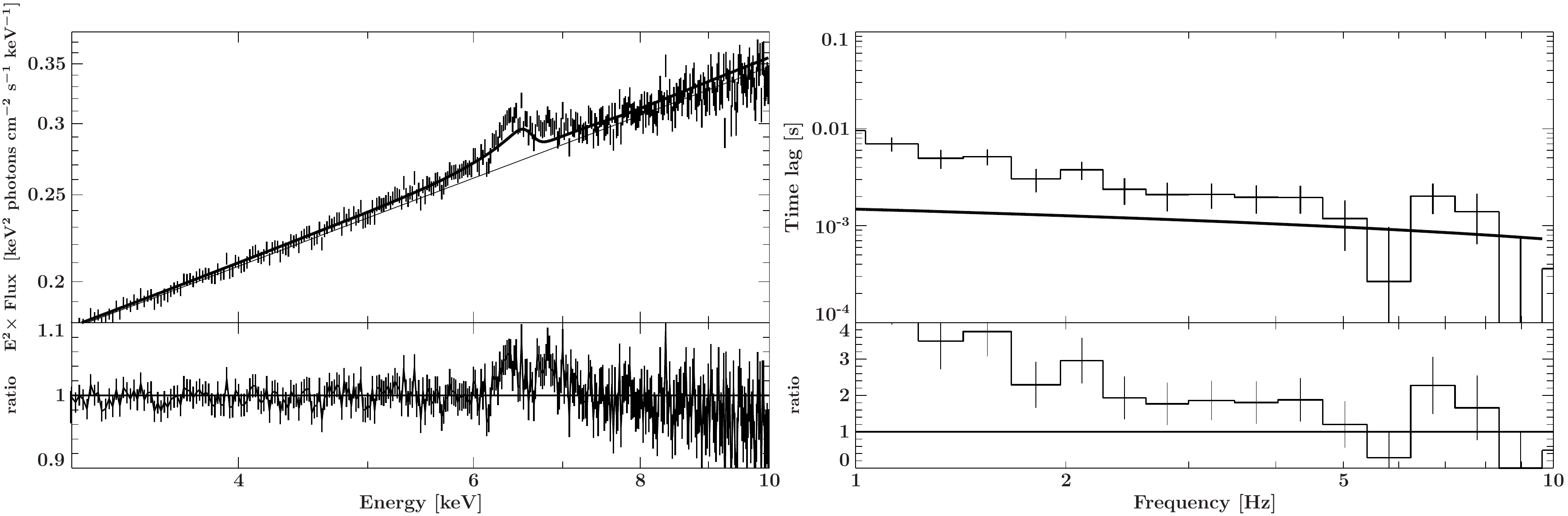}
\caption{Obs. B' (2009 spectrum, 2004 lags) spectrum (left) and lags vs frequency (right). The thick solid
lines represent the model {\sc reflags} for the best-fitting
parameters (see Table~\ref{tbl:hifreq}, right column), in the frequency
range 1--10 Hz.}
\label{fig:4+9st}                                    
\end{figure*}

However, as seen in the right panel of Fig.~\ref{fig:04}, the model is
clearly under-predicting the lags that are observed by a factor of
$\sim 2$, mainly at low
frequencies that would correspond to a light-crossing time expected
from distant reflection. Given that our focus is on the lags and our
aim is to understand whether they are compatible with being caused by
reflection, we test whether the lags are being constrained by the line shape
by adding an extra Gaussian component of variable width to the spectral
model for the 2004 data only{, that corresponds to the dataset with
higher signal-to-noise}.  In this way, we improve the fit `artificially'
so that the model can find the best-fitting contribution to the narrow 6.4~keV
core which might originate at larger radii, and thus we allow the model more 
freedom to produce larger lags which better fit the data.

\begin{table}
\centering
\begin{tabular}{lcc}
\hline \hline
Parameter      & Obs. A + extra line  & Obs. B' \\ \hline
$R_{\rm in}/R_{\rm g}$   &$ < 90 $ & $< 120$ \\
$R_{\rm out}/R_{\rm g}$ &  $(8.8^{+3.8}_{-1.6}) \times 10^3$ & $(11.0^{+2.6}_{-1.9}) \times 10^3$ \\
$H_{\rm out}/R_{\rm out}$  &  $0.84 \pm 0.04$ & $0.95^{+0.08}_{-0.14}$ \\
$H_\textrm{src}/R_{\rm g}$ & $ 430 ^{+90}_{-80}  $ & $> 380 $ \\
$\gamma$       &   $2.22^{+0.21}_{-0.33}$  & $2.62^{+0.36}_{-0.43}$ \\ 
$i$  &  $ 48.9^{+1.0}_{-0.6}  $      &  $42.3 ^{+1.3}_{-0.8}$     \\
$\Gamma$     &  $1.479 \pm 0.003$       & $1.46 \pm 0.05$ \\ 
$\log\xi$          &   $1.50^{+0.12}_{-0.24}$  & $1.65^{+0.08}_{-0.17}$ \\
$E_{\rm c}$ [{keV}] & $6.92\pm 0.02$ & $-$\\
$\sigma_{\rm c}$ [{keV}] & $0.17^{+0.03}_{-0.02}$ & $-$\\ 
$ \Omega^{\rm eff} / (2\pi)$ & 0.78 & 1.03\\ \hline
$\chi^2$ (spectrum)  &  1132         & 1335 \\
$\chi^2$ (time-lags)  &  107          &   77 \\
$\chi^2$/dof            &   1239 / 1414         & 1412 / 1417 \\
\hline
\end{tabular}
\caption{Spectral and lag fit parameters using the model
\mbox{\textsc{reflags}} plus an extra Gaussian fitted to 2004 spectrum
and lags (obs. A), and mixing the  2009 observation spectrum with the 2004
observation lags (obs. B'). 
}
\label{tbl:mix}
\end{table}

The best-fitting parameters after this procedure can be found in the first column of
Table~\ref{tbl:mix}, and an improvement in $\chi^2$ is clear.
The centroid energy for the additional emission line does not correspond to
any ionised iron fluorescence transition, hence its value may come
from a weighted mean of lines with a complex ionisation.

The decrease
in the \textsc{reflags} solid angle is compatible with the addition of the extra line, i.e.
the main model does not need to fit it anymore. Despite the attempt to artificially
improve the spectral fit to allow the model to better 
fit the large scale reflection, the lags from the model are still too low
(Fig.~\ref{fig:04pl}).

\begin{table}
\centering
\begin{tabular}{lcc}
\hline \hline
Parameter      & Obs. A + extra line  & Obs. B'\\ \hline
$R_{\rm in}/R_{\rm g}$  &$ <60  $ & $< 22$ \\
$R_{\rm out}/R_{\rm g}$ &  $(8.2^{+14.2}_{-3.6}) \times 10^3$ & $(9.1^{+12.2}_{-3.5}) \times 10^3$ \\
$H_{\rm out}/R_{\rm out}$ & $>0.06$ & $<0.05$ \\
$H_{\rm src}/R_{\rm g}$ & $  > 80  $ & $>80$ \\
$\gamma$       &   $> 1.1$  & $>1.0$ \\ 
$i$  &  $ < 31.0  $      &  $< 30.2$ \\
$\Gamma $     &  $1.459 \pm 0.003$       & $1.4398^{+0.003}_{-0.009}$ \\
$\log\xi$          &   $<1.34$  & $<1.31$\\
$E_{\rm{c}}$ [{keV}] & $6.85\pm 0.02$ & $-$\\
$\sigma_{\rm c}$ [{keV}] & $0.25\pm 0.03$ & $-$\\ 
$ \Omega^{\rm eff} / (2\pi)$ & 0.81 & 0.86 \\ \hline
$\chi^2$ (spectrum)  &  1108         &  1449 \\
$\chi^2$ (time-lags) &    74         &    78 \\
$\chi^2$/dof             &  1182 / 1402        &  1527 / 1405  \\
\hline
\end{tabular}
\caption{Spectral and lag fit parameters using the model
\mbox{\textsc{reflags}} plus an extra Gaussian (column one), and
mixing the 2009 observation spectrum with the 2004 observation lags in
the range 1--10 Hz.}
\label{tbl:hifreq}
\end{table}

Given the fact that the lags in 2004 and 2009 are very similar despite
a factor of 2 smaller error bars in 2004, we also  try to fit the 2009 spectrum
swapping 2009 and 2004 lags (B', hereinafter). This worsens the fit with respect
to fit B (Fig.~\ref{fig:4+9}, second column in
Table~\ref{tbl:mix}), as the smaller error bars in the 2004 lags push
the geometrical parameters to more extreme values, resulting in an increase of the
residuals around the iron line and a slightly larger width of the line
due to the increase of the flaring parameter accompanied by a decrease
of the outer radius. These changes also increases the
frequency at which the lag-frequency dependence becomes constant,
which corresponds to the longest time-scale of variations in the
response function).

The disc parameters in all these fits lead to a geometry where the source height,
disc flaring as well as the $H_{\rm out}/R_{\rm out}$ ratio at the outer
radius are extreme. This is due to the fact that the lags are driving the
fits towards increased outer reflection to increase their amplitude, however the spectral
model is not necessarily sensitive to small X-ray flux contributions from
outer radii or the change in iron line shape that results (since the different
line widths contributed by radii beyond a few thousand $R_{\rm g}$ cannot be resolved
by the EPIC-pn detector).

\begin{figure}
\centering
\includegraphics[width=.47\textwidth]{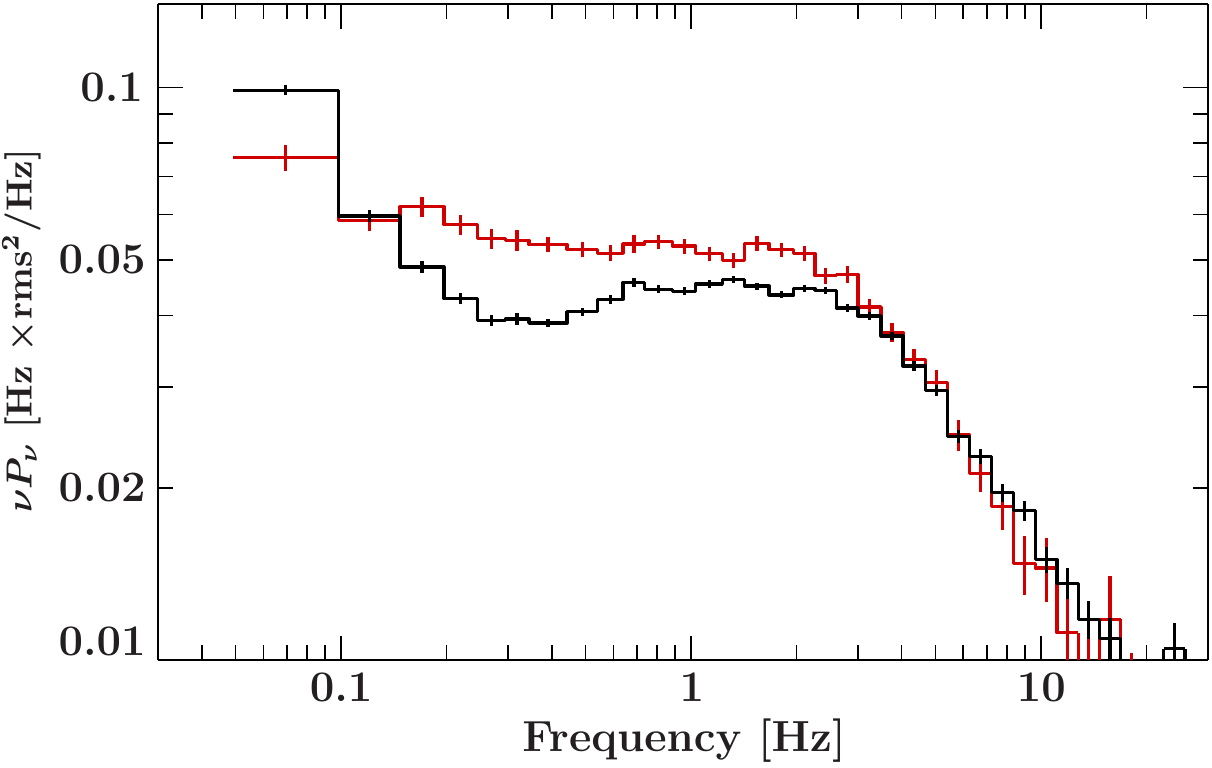}
\caption{
Power spectral densities for the 2004 (A, black) and 2009 (B, red)
observations, in the energy range 2.0 -- 10.0 keV.}
\label{fig:psds}
\end{figure}

In a scenario where the lags are caused by reflection only at higher
frequencies (e.g. with a propagation model explaining the lower-frequency lags)
the lags would correspond to reflecting regions close to
the central source of direct emission. If this is the case, excluding
the lags below a certain frequency would lead to a lower and more
plausible $H_{\rm out}/R_{\rm out}$. {
We therefore exclude lags below 1 Hz from 
the fit
{ 
(Table~\ref{tbl:hifreq})}
 }, since this frequency roughly corresponds
to the threshold frequency below which the lags relative to softer energies are consistent with
being due to propagation of disc fluctuations \citep{uttley2011}. {
In addition, this range approximately corresponds to a range in frequency of
the PSD (Fig.~\ref{fig:psds}) where the PSD has a similar shape in
both observations. }
Figs~\ref{fig:04st} and \ref{fig:4+9st} show that when lags at frequencies 
$<1$~Hz are excluded, the model cannot successfully reproduce either
the lag shape or amplitude.

\subsection{Consistency of the reflection lags model with optical/UV data}
The parameters inferred for the best-fitting reflection model, which are quite
extreme and still do not provide a good fit to the lags, can also be checked
for the implied effect on optical and UV emission from GX~339--4.  X-ray heating
of the outer disc could in principle produce a large optical/UV flux if there
is a large solid angle illuminated by the continuum, as is the case for the
geometry inferred from our lag model fits.  We can assume that each illuminated
cell in the disc absorbs a fraction of the illuminating continuum equal to $1-a(E)$,
so that the incident luminosity absorbed by each cell can be calculated for the 
best-fitting given continuum shape and model ionisation parameter.  If we make
the simplifying assumption that the absorbed luminosity dominates over any
intrinsic blackbody emission, we can equate the luminosity that is re-emitted
by the cell to the absorbed luminosity and so determine the temperature of 
blackbody radiation emitted by each cell, and hence determine the total 
reprocessed contribution to the Spectral Energy Distribution (SED).

To compare the predicted contribution to the optical/UV SED from the geometry
required by the lags model, we have extracted {\it Swift}/UVOT (bands ubb, um2, uuu, 
uvv, uw1, uw2,
uwh) spectra from a 1760 s observation of GX~339--4
made on 29 March 2009, two
days after {\it XMM-Newton} observed the source. 
\texttt{uvot2pha} was used to extract spectra for source and
background using a 6~arcsec radius, as well as extract response files. No
additional aspect correction was required.

From the best-fitting spectral parameters found in fit B' and assuming a 
high-energy cut-off at 100 keV
\citep{motta2009}, we derive an X-ray luminosity of 
$3.5 \times 10^{37}$ erg s$^{-1}$ (assuming $d = 8$ kpc; \citealt{zdziarski2004}).
This value can be used to predict the
reprocessed fluxes that are consistent with the geometries inferred from the fits, accounting for
interstellar extinction using the \textsc{xspec} model \texttt{redden}.

\begin{figure}
\centering
\includegraphics[width=0.47\textwidth]{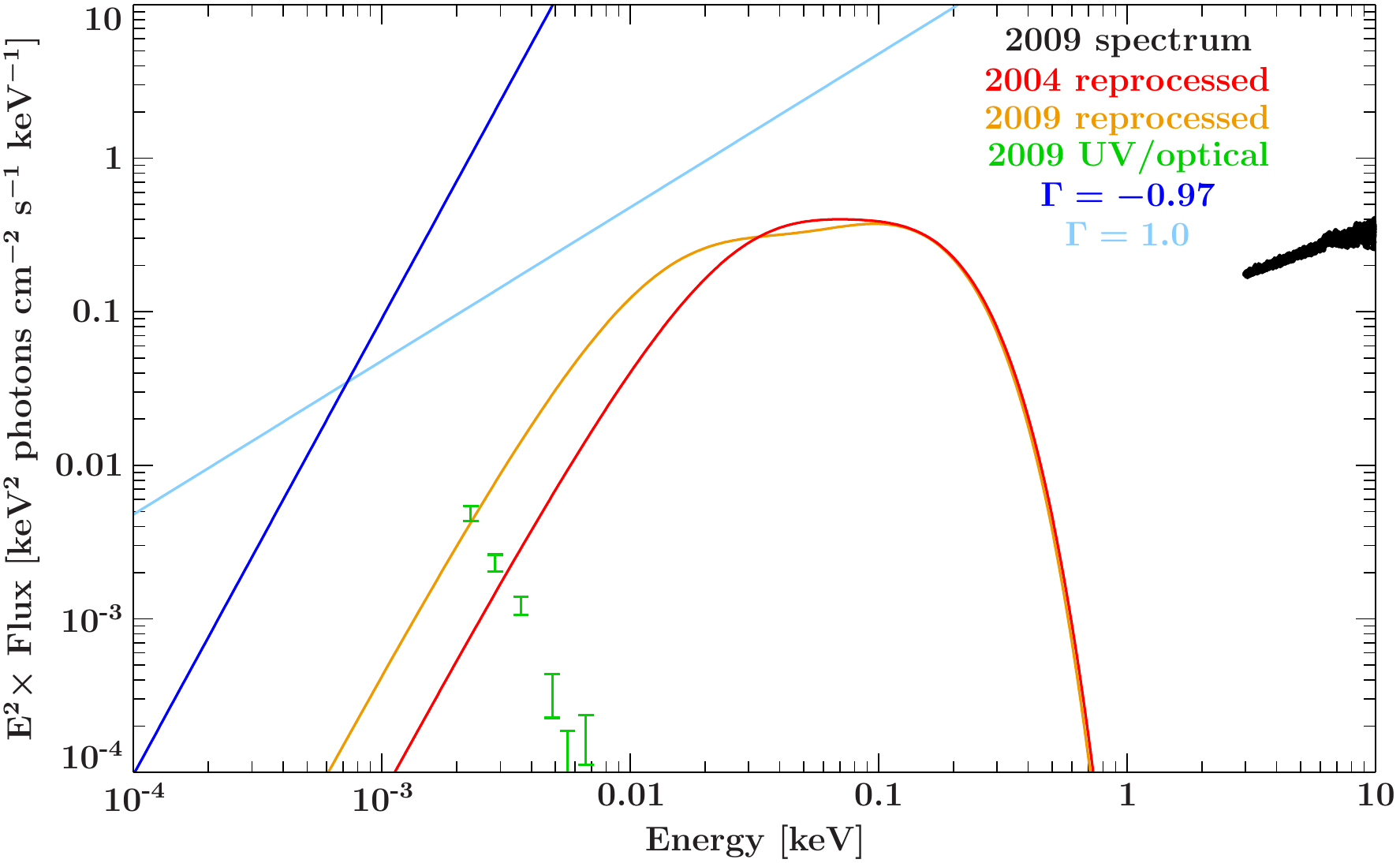}
\caption{Comparison between expected reprocessed emission using 2009
spectrum with 2004 lags (red) and 2009 lags (orange) and power-law fits to
the data with photon indices $-$0.97 (same as low energy tail of reprocessed
spectra, in {
blue}) and 1.0 (as expected from synchrotron emission coming from a
jet, in {
light blue}) after removing extinction. The data points from UVOT are shown in green.}
\label{fig:uvot}                                    
\end{figure}

Fig.~\ref{fig:uvot} shows the expected reprocessed spectra for the
two fits to the 2009 X-ray data (orange for B, red for B', black is
the 2009 X-ray spectrum). The photon index $\Gamma$ of the power-law that
characterises the reprocessed spectra at the energies covered by the
UVOT data is -0.97. In these energy ranges, dust extinction needs to
be taken into account using the multiplicative model
\texttt{redden} \citep{cardelli1989}. E(B-V) = 0.933 is the value for the extinction
calculated from IR dust maps along the line of sight towards our
source\footnote{\url{http://irsa.ipac.caltech.edu/}}.  By fitting a power-law
with a photon index of -0.97 to the
UVOT data, one finds the unabsorbed intrinsic power-law depicted in
{
blue}, which has a normalisation in the UVOT energy range that 
is several decades larger than that expected from reprocessing, and 
requires E(B-V) = 1.587 for $\chi^2$/dof=0.8/4.  Therefore the model
severely underpredicts the observed flux.
A more likely explanation for the optical/UV emission is flat-spectrum
synchrotron emission from a compact jet \citep{maitra2009}, or
magnetised hot accretion flow \citep{veledina2011}.  Assuming a power-law 
photon index of 1.0 (spectral energy index of 0), 
one also obtains a good fit 
($\chi^2$/dof=0.64/4) and a lower extinction than in the
previous case, E(B-V) = 1.089 (light blue line). This is consistent with the
results found by \citet{maitra2009} for the same object, fitting
broadband data using only synchrotron and inverse-Compton models. 
Therefore, the UVOT data cannot be explained solely by reprocessing 
in the flared disc envisaged by the lags model and is more likely to be
produced by a synchrotron process.  However, a small contribution from
reprocessing cannot be ruled out.

\section{Discussion}
\label{sec:discussion}

\subsection{Physical implications}

The analysis shown in Section~\ref{sec:results} shows that {
extreme disc geometries} are preferred for the model to reproduce the observed lags as closely
as possible within the constraints given by the overall spectral shape,
including the strength of the iron feature at 6.4 keV.
There are several effects of the reprocessing geometry on the spectral and lag data that need to
be highlighted to understand why the spectrum can be fitted well,
whereas the lags cannot. 

Firstly, while the amount of flux in the iron line
is proportional to the solid angle subtended by the disc, the ability
to determine how much of it is produced in the outer radii of the
disc (where Doppler effects are weak) 
is limited by the resolution of the {\it XMM-Newton} EPIC-pn detector.  Therefore,
the description of the geometry that could be inferred by the spectral modelling alone is
degenerate, since line emission from the largest radii (e.g. $\sim10^{5}$~$R_{\rm g}$)
cannot be resolved from that at more modest (but still large) radii (e.g. $\sim10^{4}$~$R_{\rm g}$). 
The result of this effect is that the spectral fits are not 
sensitive to variations in the outer radius of the reflector at large radii.

On the other hand, the lags are very sensitive to the geometry at large radii. 
Firstly, the lags at low frequencies increase with both the solid angle 
and light-travel time to the reflector at large radii. 
A larger outer radius and more flared disc therefore
corresponds to larger lags. However, the size-scale of the largest radius
also corresponds to a characteristic low-frequency flattening in the lag
versus frequency dependence.  This is because the frequency-dependent drop
in lags seen at higher frequencies is caused by smearing of the reflection
variability on time-scales shorter than the light-travel size-scale of
the reflector.  The reflection variability amplitude is not smeared out
for variability time-scales significantly longer than the light-travel
time to the largest disc radii, and the lag at low frequencies quickly
approaches the average light-travel delay from the reflector (diluted
by the direct continuum emission which has zero intrinsic lag). 
The frequency of this characteristic flattening in the lag-frequency
dependence is therefore a sensitive indicator of the size-scale of the
reflector.  However, it is not possible to reconcile the position of
this flattening at frequencies $\sim0.2$~Hz with the large amplitude
of the lags at low frequencies, which imply an even larger reflector
subtending an even greater solid angle to the continuum at large radii.

\begin{figure}
\centering
\includegraphics[width=.47\textwidth]{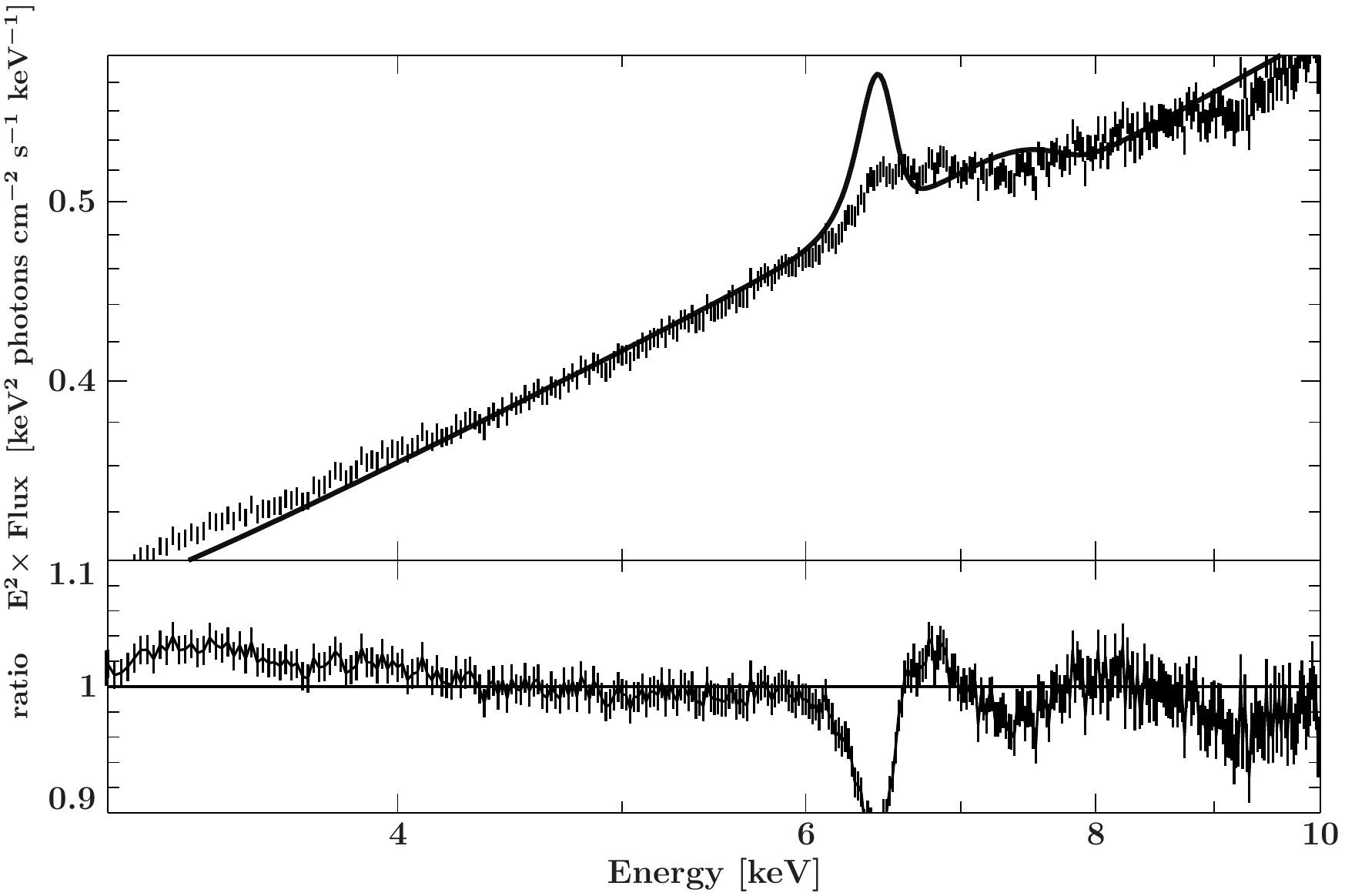}
\caption{Spectral model (\textsc{reflags}) compared to
2004 spectrum using the parameters that are derived from fitting the 2004
lags alone. The lower panel shows the ratio data/model.}
\label{fig:extreme}
\end{figure}

The result of these effects is that despite the already extreme
inferred geometries the model is clearly underpredicting the lags. 
At this point, the maximum value of the lags is now constrained by the
spectral modelling, which cannot place a tight constraint on the disc
outer radius but does limit the solid angle of the reflector.
It is instructive to consider the effects on the predicted spectral
shape when the model is fitted to the 2004 lags alone (fixing 
$\Gamma = 1.5$) and the same parameters are used to estimate the resulting
spectrum. This yields $\chi^2$/dof=66.4/16=4.15 for the fit
to the lags and yields an apparent solid angle $\Omega^{\rm eff}/2\pi = 1.51$
subtended by the disc.  The resulting spectral shape is compared to 
the data in Fig.~\ref{fig:extreme}.  Therefore, even fitting
the lag data alone with the model cannot produce a good fit, and the
fit that is obtained shows that much larger solid angles of large-scale
reflection are required than are permitted by the spectrum.  

The lags cannot be explained solely by reflection, it is therefore
necessary to invoke an additional mechanism to explain them.  
This result is perhaps not surprising, since we have previously
found evidence that in hard state BHXRBs, fluctuations in the
accretion disc blackbody emission are correlated with and precede
the variations in power-law emission \citep{uttley2011}.  Although
the disc variations seem to drive the power-law variations, this does 
not in itself explain the lags within the power-law band, which we
consider in this paper, since the disc emission only extends up to 
$\sim2$~keV.  However, as we noted in \citet{uttley2011}, at 
frequencies $<1$~Hz, the lags of the power-law emission relative to
the disc-dominated 0.5--0.9~keV band show a similar frequency-dependence
to the lags seen within the power-law band (i.e. $\tau \propto \nu^{-0.7}$). 
This strongly suggests that the lags intrinsic to the power-law are 
somehow connected to the mechanism which causes the power-law to 
lag the disc, most probably due to the propagation of accretion
fluctuations through the disc before reaching the power-law emitting hot flow.  

One possibility is that the disc is sandwiched by the hot-flow/corona
which produces power-law emission which becomes harder towards smaller
radii, leading to hard lags as fluctuations propagate inwards (e.g. 
\citealt{kotov2001,arevalo2006}). This model can explain the hard lags
in terms of propagation times in the flow, which are much larger than
light-crossing times and so can produce relatively large lags which the
reflection model struggles to produce without leading to solid angles
of large-scale reflection. Reflection may also contribute to the lags
at some level, but is not the dominant mechanism, at least at frequencies $<1$~Hz.

A more detailed analysis of the contribution of reflection to the
observed lags could be performed using datasets with higher
signal-to-noise, by e.g. searching for reflection signatures around
the iron line investigated by \citet{kotov2001}. Lags vs. energy
spectra for GX~339--4 are shown in \citet{uttley2011} for the 2004
{\it XMM-Newton} observation, and demonstrate that the current quality
of the data is not sufficient to detect these features.

{
It is also possible that multiple distinct components contribute to
the lag, e.g. associated with the different Lorentzian features that
contribute to the PSD.  This possibility could explain the apparent
`stepping' of the lag vs. frequency that appears to be linked to the
frequencies where the dominant contribution to the PSD changes from
one Lorentzian component to another \citep{nowak2000}.}

\subsection{Wider implications of combined spectral-timing models}

In this work we have considered (and ruled out) a relatively simple
model for the lags in terms of the light-travel times from a large
scale reflector.  However, it is important to stress the
generalisability of our approach to other models.  In particular, we
have shown how it is possible to combine timing and spectral
information to fit models for the geometry and spatial scale of the 
emitting regions of compact objects.  Previous approaches to use the
information from time-lags to fit models for the emitting region have
focussed on fitting the lag data (e.g. \citealt{kotov2001,poutanen2002}).
However, since these models also make predictions for the spectral
behaviour, it is possible to achieve stronger constraints on the models
by fitting the lags together with the time-averaged energy spectrum, as
done here, or with spectral-variability products such as the
frequency-resolved rms and covariance spectra \citep{revnivtsev1999,wilkinson2009,uttley2011}.  

In order to use these techniques more generally, one needs to
calculate the energy-dependent response function for the emitting
region, i.e. determine the emission as a function of time delay and 
energy.  This approach can be used to test reverberation models for 
the small soft lags seen at high frequencies in AGN 
\citep{fabian2009,zoghbi2010,demarco2011,emmanoulopoulos2011}
and BHXRBs \citep{uttley2011}, which offers the potential to map
the emitting region on scales within a few gravitational radii of the
black hole.  Other models can also be considered, e.g. to test the
propagation models for the low-frequency lags with the time-delay
expressed in terms of propagation time through the accretion flow. 
Future, large-area X-ray detectors with high time and energy-resolution,
such as the proposed {\it ATHENA} and {\it LOFT} missions, will allow
much more precise measurements of the lags in combination with good
spectral measurements, so that fitting of combined models for spectral
and timing data could become a default approach for studying the
innermost regions of compact objects. Future research in this direction
is {
strongly} encouraged.

\section*{Acknowledgments}

{
We thank the referee for their valuable comments that contributed to
the clarity of this paper.} The research leading to these results has received funding from the
European Community's Seventh Framework Programme (FP7/2007-2013) under
grant agreement number ITN 215212 ``Black Hole Universe''. 
JP was supported by the Academy of Finland grant 127512.
Based on observations obtained with {\it XMM-Newton}, an ESA science mission
with instruments and contributions directly funded by
ESA Member States and NASA.
This research has made use of the NASA/IPAC Infrared Science Archive,
which is operated by the Jet Propulsion Laboratory, California
Institute of Technology, under contract with the National Aeronautics
and Space Administration.

\label{lastpage}

\end{document}